\documentclass[fleqn,10pt]{SelfArx} 

\usepackage{lipsum} 

\definecolor{color1}{RGB}{0,0,90} 
\definecolor{color2}{RGB}{0,20,20} 

\usepackage{hyperref} 
\hypersetup{hidelinks,colorlinks,breaklinks=true,urlcolor=color2,citecolor=color1,linkcolor=color1,bookmarksopen=false,pdftitle={Title},pdfauthor={Author}}

\JournalInfo{Geomagnetism and Aeronomy, 2018, Vol. 58, No. 7} 
\Archive{DOI: 10.1134/S0016793218070046} 

\PaperTitle{
Long-Term Cyclic Variability of YZ CMi in the Context of Solar and Stellar Physics
} 

\Authors{N.\,I.~Bondar\textsuperscript{1*}, M.\,M.~Katsova\textsuperscript{2**}
} 

\affiliation{\textsuperscript{1}\textit{
Crimean Astrophysical Observatory, Nauchny, Crimea, Russia
}} 
\affiliation{\textsuperscript{2}\textit{
Sternberg State Astronomical Institute, Lomonosov Moscow State University, Moscow, Russia
}} 
\affiliation{\textsuperscript{*}e-mail: otbn@mail.ru}
\affiliation{\textsuperscript{**}e-mail: maria@sai.msu.ru}
\affiliation{Received: February 22, 2018; in final form: March 5, 2018}

\newcommand{\keywordname}{Keywords} 

\Abstract{
Manifestations of activity on the YZ CMi (M4.5Ve) flare star 
which has a rotation period of 2.77 days and belongs to a group 
of activity-saturated stars, are considered from long-term photometric 
data together with photographic measurements from 1926 to 2009. 
Long-term changes in yearly mean brightness of the star are found on a 
time scale of 27.5 years with an amplitude of $0.2-0.3^\textrm{m}$ that indicate 
variations in development of surface inhomogeneities and a large degree 
of surface spottedness in the maximum activity.

Spots are distributed unevenly over the surface, their concentration is 
higher at certain longitudes spaced by intervals equal to 0.6 phases 
of the rotation period. The recovery of active longitude (a possible 
``flip-flop'' effect) occurs over a period of about 6 years. We note 
differences in activity of this star from typical for the Sun and 
solar-like stars, which are associated with its dynamic characteristics 
$P_\textrm{cyc}$ and $P_\textrm{rot}$ and its inner structure.
}

\begin{document}

\flushbottom 

\maketitle 

\tableofcontents 

\thispagestyle{empty} 

\section{INTRODUCTION}

One of the key goals of solar and stellar physics is a study of physical 
processes on the surface of low-mass active stars within time scales 
comparable to their lifetimes. The analysis of observations in different 
spectral ranges can be used to test and develop the stellar dynamo theory. 
Thus, the study of X-ray fluxes from the coronae of active late-type stars 
on the main sequence as a function  of the stellar rotation 
(Wright et al., 2011; Reiners et al., 2014)  revealed two features in their behavior: 
fast-rotating stars having a very high level of coronal activity that 
practically is not changed when the speed of axial rotation decreases 
(stars with saturated activity) and stars with the solar-type activity 
that decreases when their rotation is decelerated  (stars with entire 
complex of the solar-type activity, including a regular cycle). 
More refined analysis (Nizamov et al., 2017) has clearly shown that the 
transition from the saturated regime, which is typical for young stars, 
to the solar-type activity occurs in G, K, and M stars at different periods 
of the axial rotation. The saturated activity regime is replaced by the 
solar-type activity when the rotation periods are 1.1, 3.3, and 7.2 days 
for stars of G2, K4, and M3 spectral classes, respectively. A detailed 
examination of G--M dwarfs reveals a phenomenological similarity of their 
activity and differences due to physical parameters, interior structure, 
magnetic field intensity, and age. 

On the whole, the level of stellar activity is characterized by such parameters 
as the flare energy, starspot area, and variations in radiation intensity in the 
optical and X-ray ranges. Baliunas et al. (1996) showed a relation between the 
observed parameters -- the rotation period of a star ($P_\textrm{rot}$) and duration 
of its cycle ($P_\textrm{cyc}$) -- and theoretical dynamo number $D$ expressed as 
$P_\textrm{cyc}/P_\textrm{rot} \sim D^{\,l}$. For F--G--K stars from the HK project with periods 
over 5 days and cycles  no longer than 20~yr, $l = 0.74$. 
This ratio allows us to choose a group of stars for which the number of 
revolutions necessary for development of a cycle is determined according to 
one law (with the same $l$).

\begin{figure*}[!th] 
\centering
\vspace*{-0.9cm}
\includegraphics[width=\linewidth]{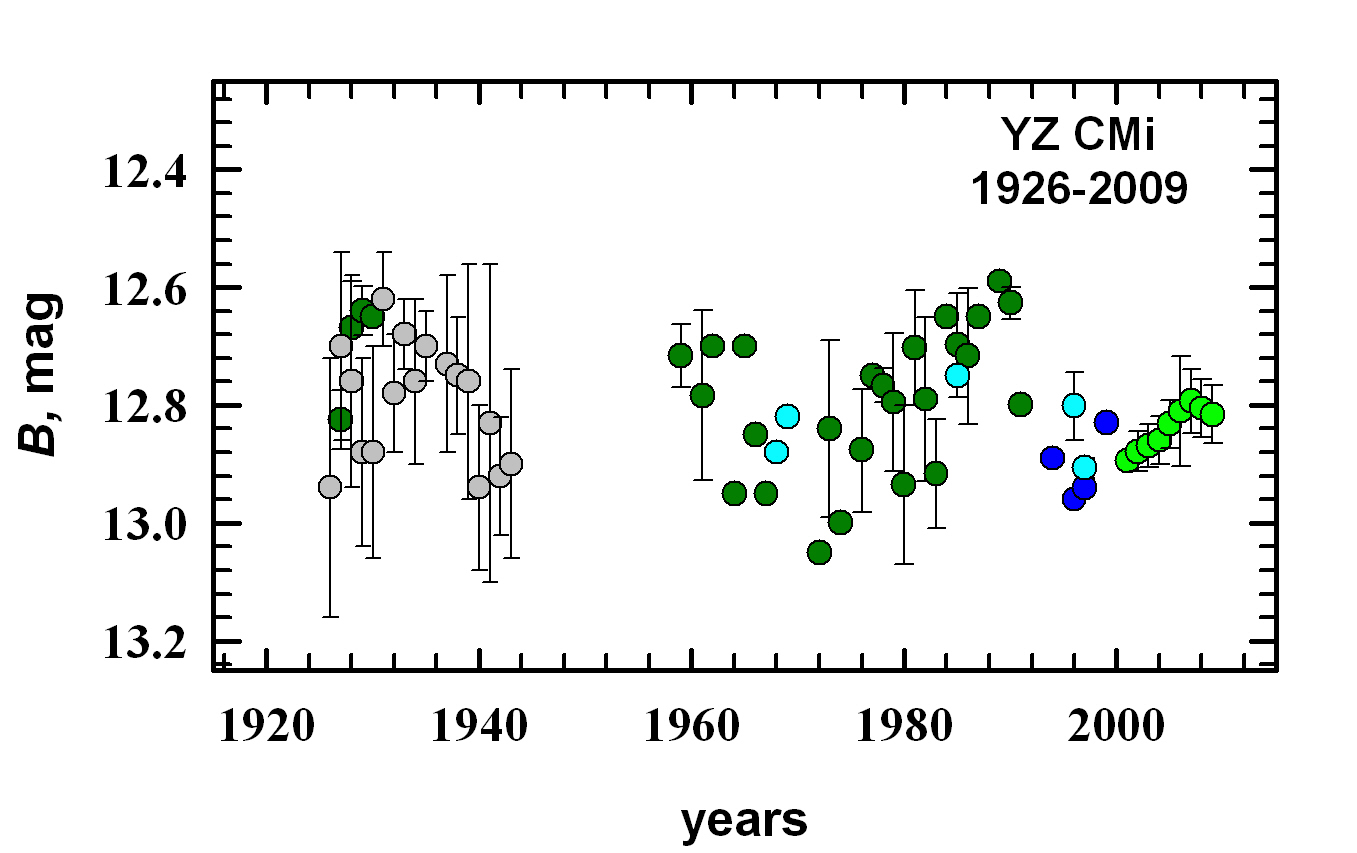}
\vspace*{0.2cm}
\caption{
Yearly mean brightness variability of YZ CMi in 1926--2009 (a) 
and a search for the activity cycle (b). Circles with different 
color represent yearly mean $B$-magnitudes: gray -- photographic data 
from Phillips and Hartmann, 1978; dark green -- Bondar, 1995, 1996; 
green -- ASAS data; cyan -- photometric data from Chugainov (1974) 
and Amado (2001); blue -- Alekseev (2001).
}
\label{Figure1}
\end{figure*}

For stars with long cycles, $l = 0.84$ (Katsova et al., 2015); 
for M-dwarfs, $l = 0.89$ (Su\'arez Mascare\~no, 2016). 
However, cycles of the considered M-dwarfs were found from the 
short time series; only some of them have several decades of available data.

In this context, observations of the long-term variability of the V833 Tau 
K-dwarf with saturated activity (Bondar, 2015; 2017) are of interest. 
This star is characterized by a very powerful corona: the coronal activity index 
$\log L_\textrm{X}/L_\textrm{bol}$ reaches --3.04. Alongside the precisely determined rotational 
modulation with a period of 1.79 days, long-term variations of optical radiation 
are revealed on a time scale of about 19 years. This indicates that in the 
saturated activity mode magnetic fields of different scales coexist and the 
process of cycle formation starts rather early when this regime has not yet been 
replaced by the solar-type activity.

\begin{figure*}[!th] 
\centering
\vspace*{-0.3cm}
\includegraphics[width=\linewidth]{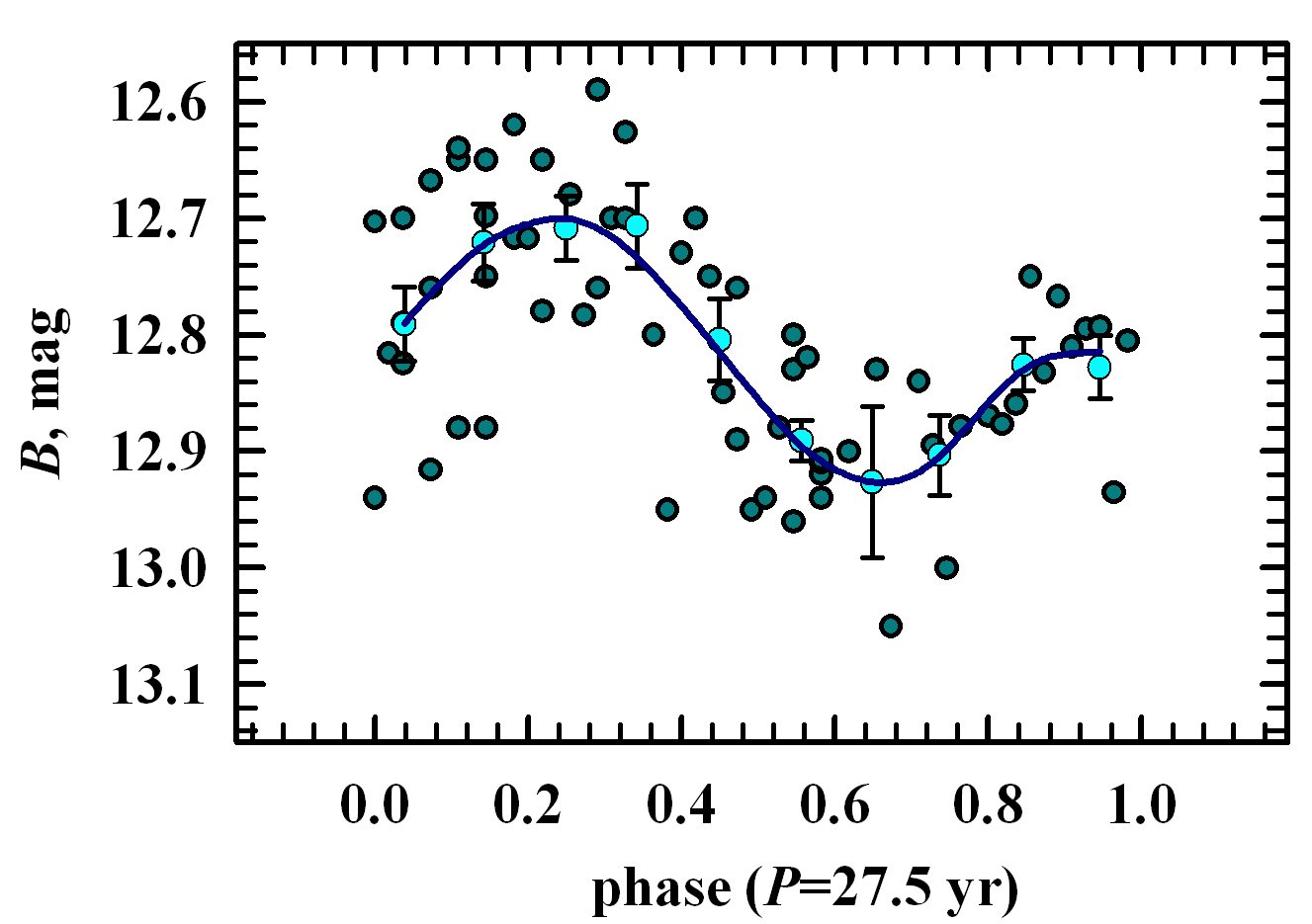}
\vspace*{0.1cm}
\caption{
Cyclic variations in yearly mean brightness of YZ CMi within a period of 27.5 years. 
Solid dark blue line marks the approximation of phase curve with the sixth-degree polynomial. 
Cyan  circles are the averaged values in bins equal to 0.1 phases, bars show mean-square errors.
}
\label{Figure2}
\end{figure*}

There is another star with saturated activity, for which long-term data series are available; 
this makes it possible to draw conclusions about the character of surface activity associated 
with starspot distribution. This star is a rapidly rotating red dwarf YZ CMi 
($V = 11.05$, M4.5 Ve, $P_\textrm{rot} = 2.77$\,days) in the solar vicinity ($d = 5.8$\,pc). 
The star is young and is undergoing the stage of arriving of a red dwarf on the main sequence 
(Eggen, 1968). Its ratio of X-ray luminosity to bolometric luminosity is evaluated as 
$\log L_\textrm{X}/L_\textrm{bol} = -2.73$. Its mass and radius parameters, $M/M_\odot = 0.31$, $R/R_\odot = 0.29$, 
indicate that it belongs to fully convective stars. The ratio $M/M_\odot = 0.35$ 
(Chabrier and Baraffe, 1997) determines the boundary between two groups of stars --
those with radiative cores and those with fully convective cores. 
Other physical properties of YZ CMi are given by Gershberg et al. (1999) 
and Bychkov et al. (2017). Recently, the studies of red dwarf activity have paid great attention 
to the late M-stars and brown dwarfs. However, detailed studies of different ranges 
of the electromagnetic spectrum cover small time intervals, that are insufficient 
for the solution of fundamental problems related to the cyclic magnetic activity of low-mass stars.

YZ CMi belongs to a small group of M-dwarfs with long-term 
data series that allow us to make conclusions about their 
activity associated with the starspot distribution.  This 
paper presents data on the frequency and energy of flares, 
surface spot area indicating a high star activity. The goal 
of this paper is to determine whether a process of starspot 
formation on this star is cyclic, to compare parameters of the 
cycle and the level of stellar activity with the activity of the 
Sun and solar-type stars. The research sets the following tasks: 
to analyze the long-term photometric data covering a time interval 
of more than 80 years, from 1926 to 2009; 
to search for the cycle and determine its parameters; 
to study the rotational modulation 
in brightness and distribution of spots at different epochs; 
to compare the level of magnetic activity of the star and 
solar-like stars by their dynamic characteristics 
$P_\textrm{cyc}$ and $P_\textrm{rot}$ associated with the dynamo number.

\section{FLARE ACTIVITY}

Development of flare activity in the YZ CMi star is detected 
at all wavelengths -- from the X-ray to radio ranges. In the optical range, 
brightness variations can reach $1^\textrm{m}-3^\textrm{m}$ with a flare duration 
of tens of minutes; less bright flares ($<0.6^\textrm{m}$) are observed 
during several minutes and even fractions of a minute; 
series of flares with different amplitudes and durations 
are often detected also. The flare amplitude increases 
towards short wavelengths. The level of flare activity of 
YZ CMi is comparable to UV Cet, a star with powerful flares 
that lent its name to the entire class of red dwarfs with flaring activity. 
No time dependence was revealed in the activity of YZ CMi. 
The flare frequency is two to three events per 10 h, but 
sometimes the interval between flares is 20 min and periods 
of quiet brightness continue for several days or even tens of days. 
YZ~CMi is one of the three stars closest to the Sun, the energy of 
powerful flares on which by several orders of magnitude exceeds 
the energy of the strongest solar flares ($E = 10^{32}$\,erg). 
These events can happen several times a year (Osten et al., 2016).

In 2008, the flare with a maximum duration of 1 hour and 
brightness variation up to 5.68$^\textrm{m}$ was registered on the 
star (Zhilyaev et al., 2011), and in 2009 the brightness 
increased by almost 6$^\textrm{m}$ during a flare that lasted for 7 h 
(Kowalski et al., 2010). The total energy of these flares was 
$\sim 10^{34} - 10^{35}$\,erg, and changes observed in the 
optical spectrum were accompanied by amplification of the 
emission in lines and in radiation in the continuum. 
This suggests that there are different mechanisms of heating, 
the effect of which varied within different phases of flares 
(Kowalski et al., 2010).

\begin{figure*}[!th] 
\centering
\vspace*{-0.9cm}
\includegraphics[width=\linewidth]{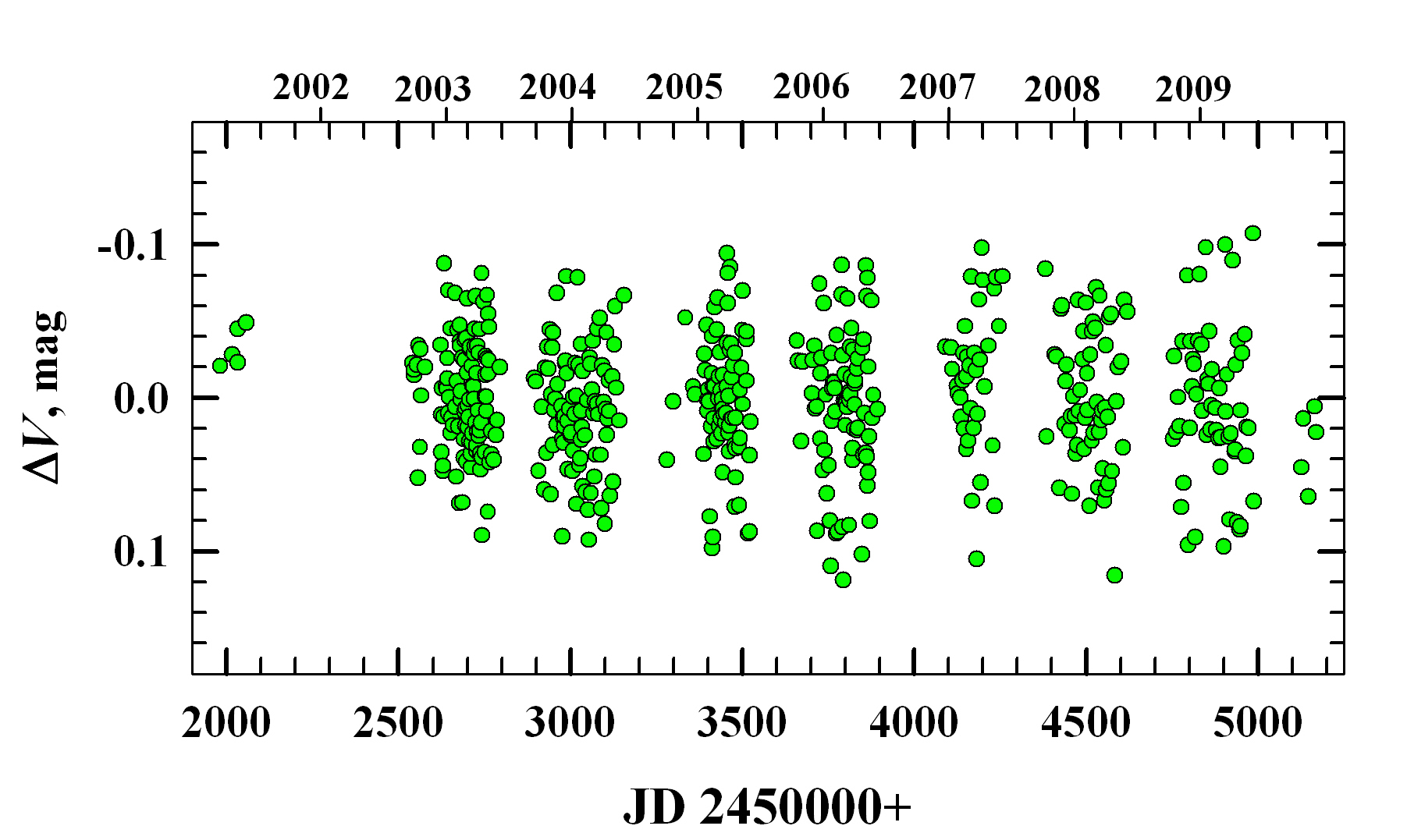}
\vspace*{0.2cm}
\caption{
Series of $\Delta V$-values for the analysis of rotational modulation according to the ASAS catalog.
}
\label{Figure3}
\end{figure*}

\section{DETECTION OF SURFACE INHOMOGENEITIES}

The kind of brightness variability of YZ CMi is typical for BY Dra-type stars. 
Active flare manifestations make it difficult to study the surface inhomogeneities, 
thereby systematic observations are required for the detection and examination of spots. 
Low-amplitude variations of YZ CMi luminosity with 2.77-day periods were discovered by 
Chugainov (1974) and confirmed by Pettersen et al. (1983). The phase light curves showed 
that the modulation of amplitude in 1972--1973 was 0.12--0.15$^\textrm{m}$, while in 1979--1980 
it was 0.18$^\textrm{m}$.  The brightness modulation with a rotation period is associated 
with the asymmetric starspot distribution.

A convincing evidence for the presence of starspots was obtained by observations 
in the IR range at a wavelength of $12\;\mu$m (Katsova and Tsikoudi, 1993) 
with the use of IRAS data. The IR excesses (differences between the flux $\lambda = 12\;\mu$m
calculated according to the surface temperature and the observed value) 
were compared with the X-ray radiation $\log(L_\textrm{X}/L_\textrm{bol})$ for 17 flare dM-stars and 14 G--K dwarfs. 
In the diagram $(m_\textrm{obs} - m_\textrm{cal}) \lambda - \log(L_\textrm{X}/L_\textrm{bol})$, 
YZ CMi and other dM-stars are located in the saturation region, where 
$\log(L_\textrm{X}/L_\textrm{bol}) = -3$, and they practically do not vary when the color excess increases. 
Among the studied stars, YZ CMi shows the greatest color excess. 
This feature indicates the presence of cool spots covering a 
considerable fraction of its surface (filling factor is close to 1). 
In models using observations from 1996--1997 (Zboril, 2003), 
cool spots with $T \sim 2800$\,K cover 10--25\%\ 
(for angles of inclination $i = 60^\circ$ and $i = 75^\circ$). 
The temperature of spots is 3000--3100 K based on the results of 
broad-band photometry (Alekseev and Bondar, 1998; Alekseev, 2001; Alekseev and Kozhevnikova, 2017); 
at certain epochs, the spots cover up to 38\%\ of the star surface, 
concentrating near the latitudes of $10^\circ - 15^\circ$.

One of the features of color characteristic of the star is that it 
becomes bluer at the epoch of higher activity. Observations by Amado et al. (2001) 
in 1996--1997 showed that color indicators $(U-B)$ and $(B-V)$ are 
changed in anti-correlation with $V$-values and color indicators in the red range. 
The star became bluer in the short wave range, when its color 
increased in the red part of the spectrum. This means that the 
active regions on the star surface include not only cool dark spots 
but facula regions also. Drawing an analogy with the known 11-year solar cycle, 
we can expect that the epochs of occurrence of large active regions 
on the surface of the star are changed by epochs of less spottedness. 
According to observations in 1926--1943 from the Harvard photographic 
collection Phillips and Hartmann (1978) suspected the cyclicity in 
spot development and estimated a cycle length of about 30 years. 
Data on measurements from the Sonneberg Observatory collection 
(Bondar, 1995; 1996), as well as photometric observations performed 
in 1994--1997 (Alekseev and Bondar, 1998), which supplement the 
photographic data, confirmed the variability in yearly mean brightness 
of the star over the indicated time span. However, due to the lack 
of observations during some years and the insufficient length 
of data series, these conclusions need further confirmation.

\begin{figure*}[!p] 
\centering
\vspace*{-0.9cm}
\includegraphics[width=0.8\linewidth]{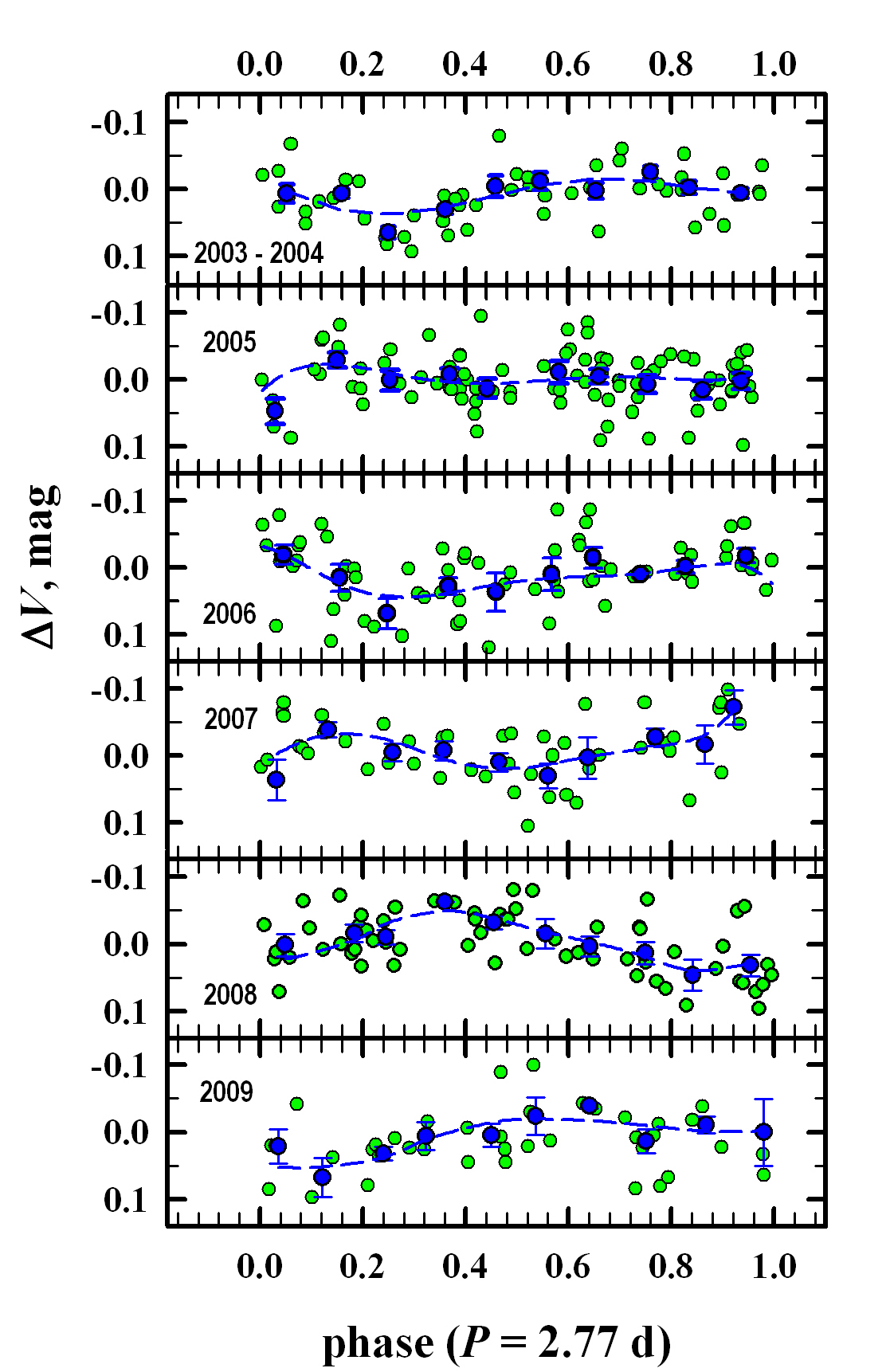}
\vspace*{0.4cm}
\caption{
Rotational modulation in brightness of YZ CMi during the seasons of 2003--2009. 
Green circles mark the observed $\Delta V$, blue circles are mean $\Delta V$-values, 
bars show their errors in bins of 0.1 phases. On each plot the dashed blue line marks 
the approximation of the phase curve with a high-degree polynomial.
}
\label{Figure4}
\end{figure*}

\section[LONG-TERM VARIATIONS IN BRIGHTNESS OF\\ YZ~CMi; A SEARCH FOR THE CYCLE]{LONG-TERM VARIATIONS IN BRIGHTNESS OF YZ~CMi; A SEARCH FOR THE CYCLE}

Estimations of the YZ CMi brightness obtained from archives of the Harvard Observatory 
and the Sonneberg Observatory correspond to the photometric $B$-values. 
In order to study the behavior of brightness of the YZ CMi luminosity over a long interval, 
we supplemented these data by the results obtained at the 
Crimean Astrophysical Observatory in 1994--1997 with the 1.25-m telescope 
(Alekseev and Bondar, 1998), and taken from the published photometric data 
(Andrews, 1968; Amado, 2001) and the ASAS photometric catalog (Pojmanski, 1997).  
Based on all these data, the yearly mean $B$-values were calculated. 
The reduction of $V$-magnitudes from the ASAS catalog was performed 
with the value of $B - V = 1.606$; the correction of $B$-values from 
Phillips and Hartmann (1978) was $-0.2^\textrm{m}$. The compiled light curve 
presents long-term variations from 1926 to 2009 (Fig. 1a).

\begin{figure*}[!th] 
\centering
\vspace*{-0.9cm}
\includegraphics[width=0.7\linewidth]{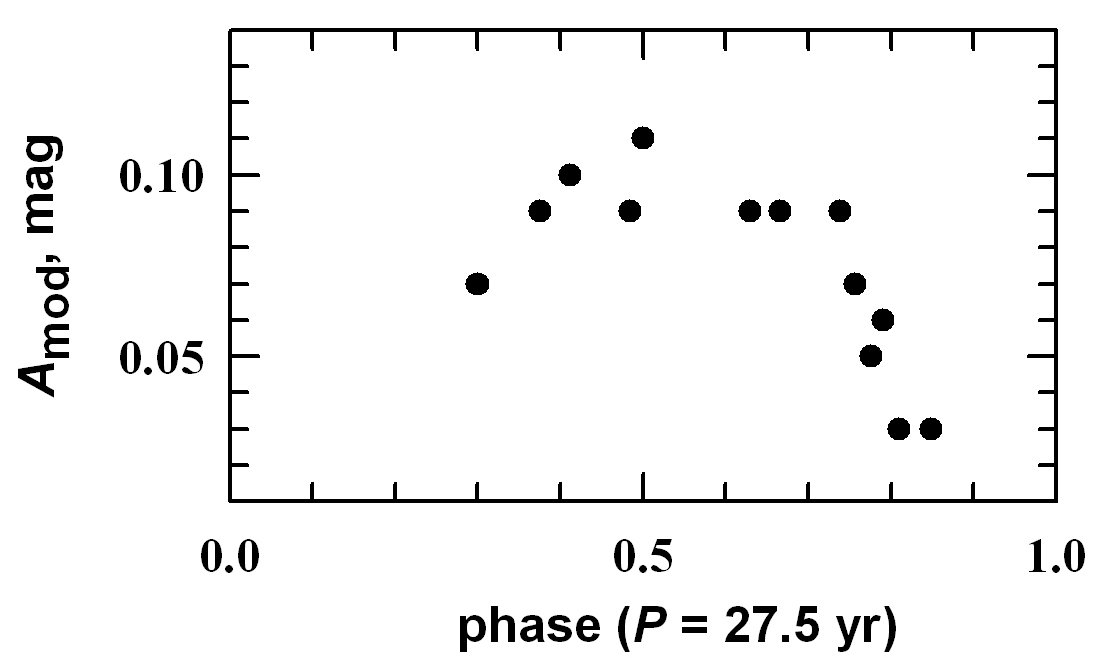}
\vspace*{0.2cm}
\caption{
Relation between the amplitude of rotational modulation and the phase of the activity cycle. 
The maximum of the activity cycle corresponds to phase 0.5.
}
\label{Figure5}
\end{figure*}

The AVE software package [http://www.gea.cesca.es] was used to 
determine the periodicity based on a generated data series containing 
65 points by the Yurkevich and Hartley methods for periods $<50$\,years. 
Figure~1b presents the periodogram obtained by the Yurkevich method, 
which shows a significant peak corresponding to a period of 27.5 years. 
Figure~2 shows the data convolution for this period. 
The large scatter of values in the maximum and minimum leaves uncertainty 
in estimation of the cycle amplitude. According to our data in the 
maximum of the light curve, i.e. without spots, $B = 12.71^\textrm{m}$, 
the same maximum level was noted by Alekseev (2001). 
The amplitude of cyclic variability is $0.2-0.3^\textrm{m}$, 
and the maximum of activity (the smallest brightness of the star) 
lasts for 4--5 years. The obtained result, $P_\textrm{cyc} = 27.5(2)$\,years, 
agrees with the 30-year cycle supposed earlier by Phillips and Hartmann (1978) 
and differs from the 10.6-year cycle by Su\'arez Mascare\~no et al. (2016). 
They found this value from the analysis of shorter row of data 
covering $\sim 9$\,years.

\section[ROTATIONAL MODULATION IN BRIGHTNESS\\ AND ACTIVE LONGITUDES]{ROTATIONAL MODULATION IN BRIGHTNESS AND ACTIVE LONGITUDES}

According to Chugainov (1974) and Pettersen et al. (1983) 
the stellar brightness varies within $0.12-0.15^\textrm{m}$ with a period of 2.77 days, 
which is equal to the period of rotation. This type of variability is produced 
by the uneven spot distribution over the surface. The amplitude of modulation 
usually does not exceed 0.1$^\textrm{m}$ (Alekseev, 2001), but it can increase up to 50\%\ 
during some epochs. We studied the rotational modulation of brightness using 
the photometric data from the ASAS-catalog for 2001--2009. Upon processing, 
excluding measurements with a large error and flares, 533 $V$-values were 
included in the data series. Cyclic changes with a period of 27.5 years 
were subtracted from the initial series as the second-degree polynomial, 
and the obtained residuals of $\Delta V$ values are used to construct phase curves 
(Fig. 3) with the indicated rotation period and an arbitrarily chosen initial epoch 
$T_0 = 2452894.9005$.

We considered phase curves for the seasons of 2003--2009 with the densest data. 
The result presented in Fig.\ 4 shows the presence of rotational modulation 
with an amplitude~\mbox{$<0.1^\textrm{m}$}. Spots cover both hemispheres and concentrate 
at active longitudes separated by $216^\circ$. The active region producing 
the highest amplitude of modulation was observed in 2003, 2006, and 2009 
at phase $\varphi = 0.25$; in 2008, the active region was amplified at phase 
$\varphi = 0.85$, while it was weakly noticeable in 2004. In 2007, 
a concentration of spots with a phase of 0.5 was observed. 
The change of active longitude within about 3 years was registered 
by the observations during the 1993--1999 seasons (Alekseev, 2001). 
However, a cycle of change of the dominating longitude is not obvious, 
and we can only suppose that it is recovered within 3--6 years. 

These results and the results published earlier have shown that the 
amplitude of rotational modulation of YZ CMi varies within the limits 
of 0.03-0.11$^\textrm{m}$, increasing  close to the maximum of activity and 
decreasing at the epoch of decreasing activity (Fig. 5). 
The data considered here do not cover all phases of the activity cycle, 
but they allow us to suggest that $A_\textrm{mod}$ changes cyclic as the 
amplitude of the cycle. This means that the unevenness of 
spot distribution increases upon cycle development, i.e. with an 
increasing of the total spottedness area; wherein one of the surfaces 
is more occupied by spots than the other one, while at minimum of 
activity spottedness of both hemispheres is almost equal and the 
rotational modulation is minimal. Calculations of starspot areas 
(Alekseev, 2001) show that at the maximum activity the area of 
starspots in one hemisphere is four times larger than in the other, 
and when the activity falls, the ratio of areas is 
$S_\textrm{max}/S_\textrm{min} \sim 1.5$.

\section{CONCLUSIONS}

YZ CMi is one of the few almost fully convective stars with a 
very strong magnetic field, about 4~kG (Reiners and Basri, 2007; 
Bychkov et al., 2017). The magnetic activity on YZ~CMi is characterized 
as flares typical for the UV~Cet-type stars and surface spots inherent to 
BY~Dra-type stars. A length of the discovered activity cycle of the star 
associated with starspot formation is 27.5 years. In order to maintain 
this cycle, the star performs tens of times more turnovers than the Sun 
or other slowly rotating G--M dwarfs with cyclic activity. The considerable 
amplitude of the cycle (more than 0.2$^\textrm{m}$) indicates  an increasing of the 
star spottedness, up to 38\%\  of its total area in the maximum of the activity 
(Alekseev and Kozhevnikova, 2017). The large area and high flare energy are 
features of difference in the activity of this star from those typical for 
the Sun and other low-mass stars. Ratio $P_\textrm{cyc}/P_\textrm{rot}$ is about $3.6 \times 10^3$, 
and the star does not belong to the G--K ensemble or early M-dwarfs, which have the similar 
relationship between $P_\textrm{cyc}/P_\textrm{rot}$ and $1/P_\textrm{rot}$. The existence of active longitudes, 
and probably their cyclic change within the range of $\sim 6$~years (manifestation 
of the ``flip-flop'' effect), is observed in the spot distribution over the 
stellar surface. This indicates the considerable role of the large-scale 
magnetic field in development of activity.

Thus, there are known two stars, V833 Tau and YZ CMi, K and M dwarfs, 
with activity saturation for which there is a long-term monitoring of 
optical variability. Being young fast-rotating objects, they possess 
very powerful coronae (the density at the corona base is tenfold higher 
than the solar value (Katsova et al., 1987), the ratio of X-ray to the 
bolometric luminosity $L_\textrm{X}/L_\textrm{bol}$ exceeds $10^{-3}$). 
These stars demonstrate a complex of non-stationary and quasi-stationary 
phenomena indicative of the presence of both large-scale and local magnetic fields. 
Their rate of axial rotation has not yet been decelerated to a level that 
would change their activity regime from the saturated to the solar type. 
However, both stars demonstrate signs of an incipient tendency towards 
more systematic and regular activity. These features of their activity 
set new tasks for observations and enhancement of theories of magnetic 
field generation in red dwarfs.

\section*{ACKNOWLEDGMENTS}

The authors are grateful to M.A.~Livshits and A.A.~Shlyapnikov for discussions. 
The study was partially supported by the Russian Foundation for Basic Research, project no. 16-02-00689.

\phantomsection
\bibliographystyle{unsrt}

\end{document}